\documentclass[aps,prd,a4paper,twocolumn,floats,nofootinbib]{revtex4} %,superscriptaddress,showpacs
\usepackage{amsmath,amssymb}
\usepackage{placeins,float}
\usepackage{color,epsfig,rotate,graphicx}
\usepackage{hyperref} 
\DeclareMathAlphabet{\mathpzc}{OT1}{pzc}{m}{it}
 
\def\beq{\begin{equation}}
\def\eeq{\end{equation}}
\def\bea{\begin{eqnarray}}
\def\eea{\end{eqnarray}}
\def\bal{\begin{align}}
\def\eal{\end{align}}
\def\bn{{\bf n}}
\def\mpc{{\rm Mpc^{-1}}}
\def\kms{~{\rm km s^{-1}}}
\def\bwt{\begin{widetext}}
\def\ewt{\end{widetext}}

\def\nn{\nonumber\\}
\def\dsum{\displaystyle\sum}

\begin{document}

%%%%%%%%%%%%%%%%%%%%%%%%%%%%%%%%%%%%%%%%%%%%%%%%%%%%%%%%%
\title{Large-scale imprint of relativistic effects in the cosmic magnification}
%%%%%%%%%%%%%%%%%%%%%%%%%%%%%%%%%%%%%%%%%%%%%%%%%%%%%%%%%
\author{Didam G.A. Duniya}%\email{adamsgwazah@gmail.com}
\affiliation{$^1$Astrophysics, Cosmology \& Gravity Centre and Department of Mathematics \& Applied Mathematics, University of Cape Town, Cape Town 7701, South Africa\\
$^2$African Institute for Mathematical Sciences (AIMS), Cape Town 7945, South Africa\\
$^3$Physics Department, University of the Western Cape, Cape Town 7535, South Africa}

%\date{\today} 

%%%%%%%%%%%%%%%%%%%%%%%%%%%%%%%%%%%%%%%%%%%%%%%%%%%%%%%%%%%%%%%%%%%%%%%%%%%%%%%%%%%%%%%%%
\begin{abstract} %\nin
Apart from the known weak gravitational lensing effect, the cosmic magnification acquires relativistic corrections owing to Doppler, integrated Sachs-Wolfe, time-delay and other (local) gravitational potential effects, respectively. These corrections grow on very large scales and high redshifts $z$, which will be the reach of forthcoming surveys. In this work, these relativistic corrections are investigated in the magnification angular power spectrum, using both (standard) noninteracting dark energy (DE), and interacting DE (IDE). It is found that for noninteracting DE, the relativistic corrections can boost the magnification large-scale power by ${\sim}40\%$ at $z=3$, and increases at lower $z$. It is also found that the IDE effect is sensitive to the relativistic corrections in the magnification power spectrum, particularly at low $z$---which will be crucial for constraints on IDE. Moreover, the results show that if relativistic corrections are not taken into account, this may lead to an incorrect estimate of the large-scale imprint of IDE in the cosmic magnification; including the relativistic corrections can enhance the true potential of the cosmic magnification as a cosmological probe.
\end{abstract} 
%%%%%%%%%%%%%%%%%%%%%%%%%%%%%%%%%%%%%%%%%%%%%%%%%%%%%%%%%%%%%%%%%%%%%%%%%%%%%%%%%%%%%%%%%

%\pacs{95.06.-g, 05.16.+a}

\maketitle

\section{Introduction}\label{sec:intro}%
The cosmic magnification \cite{Bartelmann:1999yn}--\cite{Chiu:2015tno} will be crucial in interpreting the data from future surveys that depend on the apparent flux and/or angular size of the sources, such as surveys of the 21 cm emission line of neutral hydrogen of the SKA \cite{Maartens:2015mra, Yahya:2014yva}, and the baryon acoustic oscillation surveys of BOSS \cite{Dawson:2012va, Eisenstein:2011sa}. It will be key to understanding cosmic distances, and the nature of large-scale structure in the universe. However, the fact that we observe on the lightcone, and not on a spatial hypersurface, leads to the deformation of the survey area---given that the observation angles are distorted, owing to weak (gravitational) lensing \cite{Bartelmann:1999yn, Schneider:2003yb, Schneider:2006eta, Weinberg:2012es, Yang:2013ceb}. This is the standard source of cosmic magnification in an inhomogeneous universe. However, apart from weak lensing, the area distortion is also sourced by time-delay effects \cite{Raccanelli:2013gja}.
    
Moreover, by observing on the past lightcone, the observed redshift is perturbed, by Doppler effect, which is owing to the motion of the galaxies relative to the observer, and by the gravitational potential, both local at the galaxies (i.e.~local potential effects) and also integrated along the line of sight (i.e.~integrated Sachs-Wolfe, ISW, effect). These effects surface in the cosmic magnification in redshift space---via the redshift perturbation---and together with the time-delay effect, are otherwise known as general relativistic (GR) effects. These effects are mostly known to become significant at high redshifts $z \gtrsim 1$, on very large scales. (For a range of work on GR effects in general, see~\cite{Jeong:2011as, Duniya:2015ths, Montanari:2015rga, Raccanelli:2013gja}--\cite{Renk:2016olm}.)

Forthcoming optical and radio surveys will probe increasingly large distance scales of the order of the Hubble horizon and larger, at the survey redshifts. On these cosmological scales, surveys can in principle provide the best constraints on dark energy (DE) and modified gravity models---and will be able to test general relativity itself. It is on these same scales and redshifts that the GR effects become substantial. Hence understanding the imprint of the GR effects on cosmological scales will be crucial for analysing the forthcoming data.

In this paper, the GR effects are investigated in the magnification (radial) angular power spectrum---for (standard) noninteracting DE, and for interacting DE (IDE), where DE and dark matter (DM) exchange energy and momentum, in a reciprocal manner. We start by re-deriving the standard GR magnification overdensity \cite{Jeong:2011as, Duniya:2015ths} (in first order perturbations) in Sec.~\ref{sec:DeltaMag}. In Sec.~\ref{sec:MCM} we describe a scheme for measuring the cosmic magnification (leaving out the experimental details), while in Sec.~\ref{MagPk} we discuss the magnification angular power spectrum with non-IDE. We discuss, in Sec.~\ref{IDE:Case}, the magnification angular power spectrum with IDE---with DM losing energy and momentum to DE. We conclude in Sec.~\ref{Conc}.

\section{The Relativistic Magnification Overdensity}\label{sec:DeltaMag}%
In {\it fixed-volume} surveys (with volume-limited samples), where a fixed patch of the sky is observed, the physical number of sources $N(\bn,z)$---observed in a direction $-\bn$, at a given redshift $z$ away---depends mainly on the source apparent flux $F(\bn,z)$ (or luminosity), i.e. $N(\bn,z) = N\left(F(\bn,z)\right)$. The dependence on flux invariably leads to the (de)magnification of the observed sources, given that their apparent fluxes are inherently (de)amplified in a perturbed universe. Thus a sky patch of redshift bin $dz$ and solid angle interval $d\Omega_\bn$ will contain $dN(\bn,z)$ number of magnified galaxies:
\beq\label{dN}
dN \;=\; \tilde{n}_{_F} dF \;\equiv\; \tilde{\cal N}_{\rm g} dz d\Omega_\bn,
\eeq
where $\tilde{n}_{_F}$ is the galaxy number per unit flux, measured in redshift space; $\tilde{\cal N}_{\rm g} = \tilde{n}_{_F} \tilde{\cal F}$ is the number of the (magnified) galaxies per unit solid angle per redshift bin, with $\tilde{\cal F}$ being the corresponding flux per unit solid angle per redshift bin. Moreover, we note that $\tilde{\cal F}(\bn,z)$ depends on the underlying magnification density $\tilde{\cal M}(\bn,z)$---i.e.~per unit solid angle per redshift bin. Hereafter, overbars denote background quantities, and $\delta{\tilde X} = \tilde{X} - \bar{X}$ is the perturbation in the given quantity $\tilde{X}$, with $|\delta\tilde{X}| \ll 1$.

Thus the {\it true} (observed) overdensity of magnified sources is given by \cite{Duniya:2015ths}
\beq\label{Nfrac}
\left(\dfrac{\delta\tilde{\cal N}}{\bar{\cal N}} \right)_{\rm magnified\ sources} \;=\; {\cal Q} \left(\dfrac{\delta\tilde{\cal M}}{\bar{\cal M}} \right)_{\rm magnification},
\eeq
where we have used that $\delta\tilde{\cal N}_{\rm g} = (\partial\bar{\cal N}_{\rm g}/ \partial\bar{\cal F}) \delta\tilde{\cal F}$, and by using that for magnified sources we have $\bar{\cal F} \propto \bar{\cal M}$, i.e.~the background flux per unit solid angle per redshift bin is proportional to the associated observed magnification density, it follows that $\partial\ln\bar{\cal F} = \partial\ln\bar{\cal M}$; consequently $\delta\tilde{\cal F} / \bar{\cal F} = \delta\tilde{\cal M} / \bar{\cal M}$. The quantity ${\cal Q}$ is the {\em magnification bias}~\cite{Schneider:2006eta, Jeong:2011as, Duniya:2015ths, Blain:2001yf, Kostelecky:2008iz, Schmidt:2009rh, Schmidt:2010ex, Liu:2013yna, Camera:2013fva, Umetsu:2015baa, Hildebrandt:2015kcb, Duniya:2016ibg}, given by 
\beq\label{b_M}
{\cal Q} \;\equiv\, \left. \dfrac{\partial\ln\bar{\cal N}_{\rm g}}{\partial\ln\bar{\cal F}} \right|_{\bar z},
\eeq
where $\bar{\cal N}_{\rm g} = \bar{n}_{_F} \bar{\cal F}$. (Alternatively, \eqref{Nfrac} may be obtained directly by $\delta\tilde{\cal N}_{\rm g} = (\partial\bar{\cal N}_{\rm g}/ \partial\bar{\cal M}) \delta\tilde{\cal M}$; ${\cal Q} \equiv \partial\ln\bar{\cal N}_{\rm g} / \partial\ln\bar{\cal M}$ \cite{Jeong:2011as, Duniya:2015ths}---we then proceed using $\bar{\cal F} \propto \bar{\cal M}$.) Thus we get
\beq\label{magBias}
{\cal Q} \;=\; 1 - \dfrac{5}{2} \dfrac{\partial }{\partial \bar{m}} \log_{10} \bar{n}_{_F}  \;\equiv\; 1 - \hat{s},
\eeq
with an effective slope: $\hat{s} = -\partial\ln\bar{n}_{_F}/\partial\ln\bar{\cal F}$ (see \cite{Scranton:2005ci, Zhang:2005eb, Zhang:2005pu, Bonvin:2008ni}); $m = m_* + 2.5\log_{10} ({\cal\tilde F_*/ \tilde F})$ is the apparent magnitude, and $m_*$ is the apparent magnitude at the (fixed) initial value $\tilde{\cal F}_*$ of the flux density. Note that throughout this work we assume surveys which are independent of the source apparent angular size (but see \cite{Schmidt:2009rh, Schmidt:2010ex, Liu:2013yna, Duncan:2016kko, Duniya:2016ibg} for size-dependent analysis). 

Thus the {\em observed} magnification density perturbation \eqref{Nfrac}, is given by $\Delta^{\rm obs}_{_{\cal M}} (\tilde{\cal M}) \equiv \delta\tilde{\cal N}_{\rm g} (\tilde{\cal M}) / \bar{\cal N}_{\rm g}$---which is automatically {\em gauge-invariant} (given that it is an observable):
\beq\label{Delta:obs} 
\Delta^{\rm obs}_{_{\cal M}} (\bn,z)  \;=\; {\cal Q}(z)\, \tilde{\delta}_{_{\cal M}}(\bn,z),
\eeq
where $\tilde{\delta}_{_{\cal M}} \equiv \delta\tilde{\cal M}/\bar{\cal M}$ is the magnification density contrast. Obviously, by \eqref{magBias} (see also \cite{Camera:2013fva, Scranton:2005ci, Zhang:2005eb, Zhang:2005pu, Bonvin:2008ni}) the magnification bias exists only if $\hat{s} \neq 1$. Moreover, provided the background number density $\bar{n}_{_F}$ varies with redshift (or magnitude), the magnification bias cannot be unity. Thus for ${\cal Q}=1$, it implies that $\bar{n}_{_F}$ = constant.

In the presence of magnification, the transverse area per unit solid angle---in redshift space---$\mathpzc{\tilde{A}}$ becomes distorted by a factor $\mu = \tilde{\cal M}/\bar{\cal M}$, given by
\beq\label{Magnfcn}
\mu^{-1} \;\equiv\; \dfrac{\mathpzc{\tilde A}}{\mathpzc{\bar A}} \;=\; \dfrac{\tilde{D}^2_A}{\bar{D}^2_A}, 
\eeq 
where $\tilde{D}_A$ is the associated angular diameter distance to the source. Thus an overdense, inhomogeneous region will have a magnification factor $\mu > 1$ (objects appear closer than they actually are, and the screen-space area appears reduced), and an underdense region will have $\mu <1$ (objects appear farther, and the screen-space area appears enlarged), while a smooth, homogeneous region will have $\mu = 1$ (objects are seen at their true position, with the screen-space area remaining unchanged). Moreover, for ($\mu < 1$) $\mu > 1$ the observed flux is (de)amplified; for $\mu = 1$ the observed flux is equal to the true flux.

The area density is usually given as $\mathpzc{\tilde{A}} = \mathpzc{\tilde{A}}(>\tilde{\cal F})$ and $\mathpzc{\bar{A}} = \mathpzc{\bar{A}}(>\tilde{\cal F}/\mu)$---corresponding to sources with flux density greater than $\tilde{\cal F}$ and $\tilde{\cal F}/\mu$, respectively. Note that given $\bar{\cal F} {\propto}\/ \bar{\cal M}$, it follows that $\bar{\cal F} = \tilde{\cal F}/\mu$ (up to first order).

\subsection{The transverse area density}%
We compute the screen-space area density---i.e.~the area per unit solid angle transverse to the line of sight---in redshift space. The transverse area element is%
\beq\label{AngDist}
dA \;=\; \mathpzc{\tilde{A}}(\bn,z) d\Omega_\bn \;=\; \tilde{D}^2_A(\bn,z) d\Omega_\bn,
\eeq 
where the area density $\mathpzc{\tilde{A}}$ is in a given redshift bin. In real coordinates $\tilde{x}^\alpha$, we have
\bea\label{dA1}
dA &\;=\;& \sqrt{-\tilde{g}}\, \epsilon_{\mu\nu\alpha\beta}\, \tilde{u}^\mu \tilde{\ell}^\nu d\tilde{x}^\alpha d\tilde{x}^\beta,\\ \label{dA2}
&\;\equiv\; & \mathpzc{A} (\theta_{_O},\vartheta_{_O})\, d\theta_{_O} d\vartheta_{_O}, 
\eea
which is evaluated at a fixed $z$, with $\theta_{_O}$ and $\vartheta_{_O}$ being the zenith and the azimuthal angles, respectively, at the observer $O$; $\tilde{u}^\nu$ is the 4-velocity of the observer. The 4-vector $\tilde{\ell}^\nu$ is orthogonal to the line of sight, i.e.~$\tilde{u}_\nu\tilde{\ell}^\nu=0$, with its background part being purely spatial, where~\citep{Jeong:2011as}
\beq\label{ell:defn}
\tilde{\ell}^\nu \;=\; \tilde{u}^\nu + \dfrac{\tilde{n}^\nu}{\tilde{n}^\alpha \tilde{u}_\alpha},
\eeq
where $\tilde{n}^\nu = d\tilde{x}^\nu / d\lambda$ is a tangent 4-vector to the photon geodesic $\tilde{x}^\nu(\lambda)$, with $\lambda$ being an affine parameter. 

Note that $\mathpzc{\tilde{A}}$ and $\mathpzc{A}$ are the area densities in redshift space and in real space, respectively. From \eqref{dA2}, we have (henceforth assuming flat space)
\beq\label{A:defn}
\mathpzc{A} \;=\; \sqrt{-\tilde{g}}\, \epsilon_{\mu\nu\alpha\beta}\, \tilde{u}^\mu \tilde{\ell}^\nu \dfrac{\partial\tilde{x}^\alpha}{\partial\theta_{_S}} \dfrac{\partial\tilde{x}^\beta}{\partial\vartheta_{_S}} \left| \left. \dfrac{\partial(\theta_{_S},\vartheta_{_S})}{\partial(\theta_{_O},\vartheta_{_O})} \right| \right. ,
\eeq
where $\tilde{g} = {\rm det}(\tilde{g}_{\mu\nu})$, with $\theta_{_S} = \theta_{_O} + \delta\theta$ and $\vartheta_{_S} =~\vartheta_{_O}~+~\delta\vartheta$ being the angles at the source $S$. Thus after some calculations (see Appendix \ref{MagDistortn}), we obtain
\bea\label{A:exprssn}
\mathpzc{A} &\;=\;& \mathpzc{\bar A}\left[1 - 3D - \phi + \bar{n}^i B_{|i} - \dfrac{1}{2} \delta{g}_{\alpha\beta}\bar{n}^\alpha \bar{n}^\beta \right. \nonumber \\
&& \quad\quad + \left. 2\dfrac{\delta{r}}{\bar{r}} + \left(\cot{\theta} +\partial_{\theta}\right)\delta{\theta} +\partial_{\vartheta}\delta{\vartheta}\right],
\eea
where $\mathpzc{\bar A}(\bar{z}) = a(\bar{z})^2\bar{r}(\bar{z})^2\sin{\theta}$ is the background area density---computed at $\bar{z}$, in the unperturbed universe, with $r = \bar{r} + \delta{r}$ being the comoving radial distance. The parameters $B$, $D$ and $\phi$ are scalar metric potentials.

\subsection{The magnification distortion}%
Here we compute the fractional perturbation $\tilde{\delta}_{_{\cal M}}$ in the magnification density $\tilde{\cal M}$. By~\eqref{Magnfcn}, we have $\mu^{-1} = \mathpzc{\tilde A} / \mathpzc{\bar A} = 1 + \tilde{\delta}_{\mathpzc{A}}$, where $\tilde{\delta}_{\mathpzc{A}} \equiv \delta\mathpzc{\tilde{A}} / \mathpzc{\bar A}$ is the redshift-space area density contrast. Hence by taking a gauge transformation, from real to redshift space, we have
\beq\label{perp_dA}
\tilde{\delta}_{\mathpzc{A}}(\bn,z) \;=\; \delta_{\mathpzc{A}}(\bn,z) -\dfrac{d\ln\mathpzc{\bar{A}}}{d\bar{z}} \delta{z}(\bn,z),
\eeq
where $\delta_{\mathpzc{A}} \equiv \delta\mathpzc{A} / \mathpzc{\bar A}$ is the real-space area density contrast, with $\mathpzc{\bar A}$ remaining the same for both $\mathpzc{A}$ and $\mathpzc{\tilde A}$. In \eqref{perp_dA}, we used that the conformal time perturbation $\delta{\eta} = (\partial\bar{\eta} / \partial\bar{z}) \delta{z}$; $\delta{z}=z-\bar{z}$ is the redshift perturbation. 

Thus given~\eqref{A:exprssn} and~\eqref{perp_dA}, we obtain
\bea\label{Mag2}
\mu^{-1} &=& 1 - 3D - \phi + \bar{n}^i B_{|i} - \dfrac{1}{2} \delta{g}_{\alpha\beta}\bar{n}^\alpha \bar{n}^\beta + 2\dfrac{\delta{r}}{\bar r}  \nonumber\\
&+& \left(\cot{\theta} +\partial_{\theta}\right)\delta{\theta} +\partial_{\vartheta}\delta{\vartheta} + 2a\left(1 - \dfrac{1}{\bar{r} {\cal H}}\right)\delta{z},\quad
\eea
where ${\cal H} = a' /a$ is the comoving Hubble parameter, with a prime denoting differentiation with respect to conformal time $\eta$, $a=(1+\bar{z})^{-1}$ being the scale factor, and
\beq
\dfrac{d\mathpzc{\bar A}}{d\bar{z}} \;=\; -2a \left(1 - \dfrac{1}{\bar{r} {\cal H}}\right) \mathpzc{\bar A} .
\eeq
After some calculations (see Appendix \ref{MagDistortn}), given \eqref{Mag2} and $\mu^{-1}=1-\tilde{\delta}_{_{\cal M}}$, we obtain the relativistic magnification distortion as
\bea\label{MagDens}
\tilde{\delta}_{_{\cal M}}(\bn,z) &=& -\int^{\bar{r}_{_S} }_0{ d\bar{r}\left(\bar{r} - \bar{r}_{_S} \right) \dfrac{\bar{r}}{\bar{r}_{_S} } \nabla^2_\perp \left(\Phi + \Psi\right) (\bn,z)} \nn
&& +\; 2\Psi(\bn,z) - \dfrac{2}{\bar{r}_{_S}} \int^{\bar{r}_{_S}}_0{ d\bar{r} \left(\Phi + \Psi\right) (\bn,z)} \nn
&& +\; 2\left(1 -\dfrac{1}{\bar{r}_{_S} {\cal H}(\bar{z})}\right) \Big[ \Phi(\bn,z) + V_\parallel(\bn,z) \nn
&&\hspace{1.1cm} - \int^{\bar{r}_{_S} }_0{d\bar{r} \left(\Phi' + \Psi' \right) (\bn,z)} \Big] ,
\eea
where $\bar{r}_{_S} = \bar{r}(\bar{z}_{_S})$ is the background comoving distance at $S$, $\Phi$ and $\Psi$ are the Bardeen potentials, with $V_\parallel \equiv \bn \cdot {\bf V} =\bar{n}^i\partial_iV$ being the velocity component along the line of sight, and $V$ is a gauge-invariant velocity potential; see Appendix \ref{MagDistortn}, i.e.~\eqref{InvPhi}--\eqref{InvVel}. (Note that nonintegral terms in \eqref{MagDens} denote the {\it relative} values, those at $S$ relative those at $O$, accordingly.) The squared operator $\nabla^2_\perp = \nabla^2 - (\bar{n}^i\partial_i)^2 + 2\bar{r}^{-1} \bar{n}^i\partial_i$ is the Laplacian on the screen space---transverse to the line of sight (the various terms retaining their standard notations). In \eqref{MagDens}, the first line gives the weak lensing term; the remaining lines together give the GR corrections.

Thus given \eqref{MagDens}, we rewrite the (observed) relativistic magnification overdensity \eqref{Delta:obs} (see also \cite{Jeong:2011as, Duniya:2015ths, Bonvin:2008ni}):
\beq\label{Delta:obs2}
\Delta^{\rm obs}_{_{\cal M}} (\bn,z) \;=\; \Delta^{\rm std}_{_{\cal M}}(\bn,z) \;+\; \Delta^{\rm GR}_{_{\cal M}}(\bn,z),
\eeq
where the weak lensing magnification is taken as the {\em standard} term, given by 
\beq\label{MagStd} 
\Delta^{\rm std}_{_{\cal M}} \;\equiv\; -{\cal Q} \int^{\bar{r}_{_S} }_0{ d\bar{r}\left(\bar{r} - \bar{r}_{_S} \right) \dfrac{\bar{r}}{\bar{r}_{_S}} \nabla^2_\perp \left(\Phi + \Psi\right)},
\eeq
and the GR corrections are given by 
\bea\label{MagGR}
\Delta^{\rm GR}_{_{\cal M}} &\equiv & 2{\cal Q} \left\lbrace  \left(1 -\dfrac{1}{\bar{r}_{_S} {\cal H}}\right) \left[ V_\parallel  - \int^{\bar{r}_{_S}}_0{d\bar{r} \left(\Phi' + \Psi' \right) } \right] \right. \nn
&+& \Psi + \left. \left(1 -\dfrac{1}{\bar{r}_{_S} {\cal H}}\right) \Phi - \dfrac{1}{\bar{r}_{_S}} \int^{\bar{r}_{_S}}_0{ d\bar{r} \left(\Phi + \Psi\right)} \right\rbrace .\quad
\eea
It should be noted that magnification of sources is only one of the effects (along with cosmic shear~\cite{Weinberg:2012es, Umetsu:2015baa, Bonvin:2008ni, VanWaerbeke:2009fb, Duncan:2013haa, Gillis:2015caa}) of weak lensing. However, weak lensing is not the only cause of cosmic magnification; other causes include \eqref{MagGR}: the Doppler effect (first term in square brackets), which is sourced by the line-of-sight relative velocity between the source and the observer; the ISW effect (second term in square brackets)---sourced by the integral of the time variation of the gravitational potentials; the time-delay effect (last integral term), and the source-observer relative gravitational potential effects (nonintegral potential terms). For example, when a source is moving towards the observer its flux becomes magnified: this is Doppler effect, i.e.~{\em Doppler magnification} (also referred to as ``Doppler lensing''~\cite{Bacon:2014uja, Raccanelli:2016avd}). Time delay also causes magnification by broadening the observed flux. Moreover, if the gravitational potential well (i.e.~the potential difference) between the source and the observer is deep enough it can also result in flux magnification, specifically when the source is at the potential crest with the observer at the trough---e.g.~sources with sufficiently lower masses relative to our galaxy (the Milky Way): signals from such sources reaching an observer on earth will appear magnified (when other effects are insignificant).

\section{Measuring the Cosmic Magnification}\label{sec:MCM}
A generic sample of cosmic objects in the sky would inherently contain both an ``unmagnified'' fraction and a ``magnified'' fraction (see e.g. \cite{Jeong:2011as, Duniya:2015ths, Bonvin:2008ni, Duniya:2016ibg, Challinor:2011bk, Camera:2014bwa}), where the magnified fraction is proportional to the magnification bias. However, during observations all events are measured together without any distinctions of these fractions---only the number density, i.e.~number of objects per unit solid angle per redshift bin, is measured. Nevertheless, the unmagnified fraction is volume dependent, while the magnified fraction is flux (or luminosity) dependent \cite{Duniya:2016ibg}. Thus, in order to measure solely the magnified fraction, i.e. the magnification overdensity, the observation is done on a fix-sized survey volume.

Observers sometimes split the survey sample into magnitude bins $\Delta{m}$, i.e.~instead of redshift bins $\Delta{z}$; thus compute the galaxy number per unit solid angle in a given $\Delta{m}$---which is essentially $\tilde{\cal N}_{\rm g}$. By noting the magnification factor $\mu = \tilde{\cal M} / \bar{\cal M}$, then \eqref{Nfrac} and \eqref{magBias} yield the following scheme (here we leave out the experiment details, but see e.g.~\cite{Gillis:2015caa}):
\beq\label{mu_i}
\mu_i \;=\; 1 \;+\; \dfrac{\tilde{\cal N}_{\rm g}(m_i) - \bar{\cal N}_{\rm g}(m_i)}{\left(1-\hat{s}(m_i)\right)\bar{\cal N}_{\rm g}(m_i)},
\eeq
where $\mu_i = \mu(m_i)$ are the values for galaxies with magnitudes $m_i=m_1,\, m_2,\, m_3,\,\cdots$, in a given $\Delta{m}$. Obviously, we have $|\mu -1| \ll 1$ (i.e.~at first order perturbation). 

In order to optimally estimate $\mu$, a weighting scheme is crucial---each $\mu_i$ is associated with a certain weighting function $w_i=w(m_i)$ (see e.g.~\cite{Gillis:2015caa, Menard:2002vz, Scranton:2005ci}), which may be thought of as a ``probability distribution function'' in the given magnitude (or redshift) bin. Thus the effective estimator for each bin, is given by \cite{Gillis:2015caa} 
\beq\label{Totmu}
\hat{\mu} \;=\; \dfrac{\sum_i{w_i\mu_i} }{\sum_i{w_i}}, 
\eeq 
with the associated standard error given by
\beq\label{stdError}
\sigma_{\hat{\mu}} \;\approx\; \left(\dsum_i{w_i}\right)^{-\frac{1}{2}},
\eeq
where the given error is only a simplistic (illustrative) approximation; a more rigorous approach may be necessary. Thus in the case where $\Delta{m} \to 0$, i.e.~infinitesimally small, the summations transform to integrals over $dm$. It should be noted that any survey that can measure magnification can also measure shear (see e.g.~\cite{Gillis:2015caa}). Moreover, the true (physical) magnification effect on cosmic objects is quantified by ${\cal Q}\, \tilde{\delta}_{_{\cal M}} = (1-\hat{s})(\mu - 1)$, i.e.~at first order perturbations. (In fact, the method given by \cite{Heavens:2011ei} can also be applied to isolate the magnification overdensity in the GR density perturbation \cite{Jeong:2011as, Duniya:2015ths, Duniya:2016ibg, Challinor:2011bk, Yoo:2010ni, Bonvin:2011bg, Yoo:2014kpa, Alonso:2015uua}.)

\section{The Magnification Angular Power Spectrum}\label{MagPk}%
The magnification overdensity \eqref{Delta:obs2} may be expanded in spherical multipoles, given by
\bea\label{Delta:ell}
\Delta^{\rm obs}_{_{\cal M}} (\bn,z) &=& \dsum_{\ell m} {a_{\ell m}(z) Y_{\ell m}(\bn)}, \nonumber \\
 a_{\ell m}(z) &=& \int{ d^2\bn\, Y^*_{\ell m}(\bn) \Delta^{\rm obs}_{_{\cal M}} (\bn,z) },
\eea
where $Y_{\ell m}(\bn)$ are the spherical harmonics and $a_{\ell m}$ are the multipole expansion coefficients, with the asterisk denoting complex conjugate. The angular power spectrum observed at a source $z_{_S}$ may then be computed as follows:
\bea\nonumber %\label{Cl_TT1}
C_\ell(z_{_S}) &\;=\;& \left\langle{\left.\left| a_{\ell m}(z_{_S}) \right.\right|^2 }\right\rangle, \\ \label{Cl_TT2}
&\;=\;& \dfrac{4}{\pi^2} \int{dk\, k^2 \Big|f_\ell(k,z_{_S}) \Big|^2 },
\eea
where by using the transformation to spherical harmonics (see \cite{Bonvin:2011bg}), we have
\bwt
\begin{align}\label{f_ell}
f_\ell(k,z_{_S}) \;=&\;\; 2{\cal Q}(z_{_S}) \left\lbrace  j_\ell(k\bar{r}_{_S}) \Phi(k,z_{_S}) - \dfrac{1}{\bar{r}_{_S}} \int^{\bar{r}_S}_0{d\bar{r}\, j_\ell(k\bar{r}) \left[2 - \dfrac{(\bar{r}-\bar{r}_{_S})}{\bar{r}} \ell(\ell+1) \right] \Phi(k,\bar{r})} \right. \nn
&\quad +\left. \left(1 - \dfrac{1}{\bar{r}_{_S} {\cal H}}\right) \left[ j'_\ell(k\bar{r}_{_S}) V^\parallel_m(k,z_{_S}) +  j_\ell(k\bar{r}_{_S}) \Phi(k,z_{_S}) - 2\int^{\bar{r}_S}_0{d\bar{r}\, j_\ell(k\bar{r}) \Phi'(k,\bar{r})}\right] \right\rbrace ,
\end{align}
\ewt
where $V^\parallel_m$ is the line-of-sight matter peculiar velocity (i.e. relative to the observer); $j'_\ell(k\bar{r}) = \partial j_\ell(k\bar{r}) / \partial(k\bar{r})$, and $j_\ell$ is the spherical Bessel function. Henceforth, we use the conformal Newtonian metric---with $\Psi=\Phi$. 

\begin{figure}\centering
\includegraphics[scale=0.4]{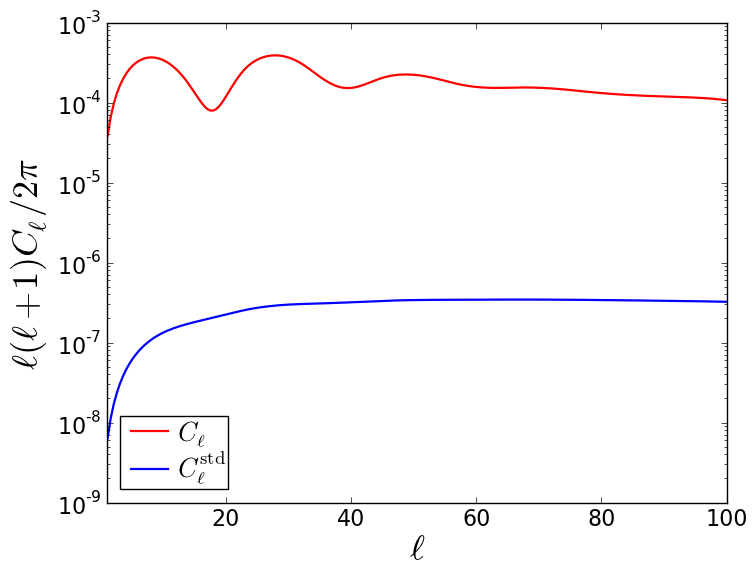}
\caption{The magnification (radial) angular power spectrum at $z_{_S}=0.1$ with ${\cal Q}=1$, for noninteracting DE scenario. The red line is the full power spectrum $C_\ell$ with all GR corrections included (i.e.~for $\Delta^{\rm obs}_{_{\cal M}}$, given by \eqref{Delta:obs2}), while the blue line is the standard power spectrum $C^{\rm std}_\ell$ containing only the weak lensing effect (i.e.~for $\Delta^{\rm std}_{_{\cal M}}$, given by \eqref{MagStd}).}\label{fig:1}  %These results agree with the work by \cite{Bonvin:2011bg}
\end{figure} 

By adopting the matter density parameter $\Omega_{m0}=0.24$ and Hubble constant $H_0=73~\kms\cdot\mpc$, we compute the (radial) magnification angular power spectrum \eqref{Cl_TT2} in the late-time universe. Firstly, we compute the angular power spectrum for a standard, noninteracting DE scenario (in this section)---assuming cosmic domination by DE and matter (dark plus baryonic); then for an IDE scenario (in Sec. \ref{IDE:Case}). We use (Gaussian) adiabatic initial conditions (see \cite{Duniya:2015ths, Duniya:2013eta, Duniya:2015nva, Duniya:2015dpa}) for the perturbations, in noninteracting DE and in IDE, accordingly. Throughout this work, we initialize evolutions at the decoupling epoch, $1+z_d = 10^3 = a^{-1}_d$. 

We take DE as a fluid with a parametrized equation of state parameter, given by \cite{Chevallier:2000qy, Linder:2002et}
\beq\label{CPL:EoS}
w_x(a) \;=\; w_0 +w_a(1-a),
\eeq
where we choose the (free) constants $w_0=-0.8$ and $w_a=-0.2$. Henceforth, we adopt a DE physical sound speed $c_x=1$ and a magnification bias ${\cal Q}=1$, for all numerical computations. (Throughout this work, the DE equation of state parameter $w_x$ is used as given by \eqref{CPL:EoS}.) Note that given our consideration of $C_\ell$, which is evaluated at a fixed $z$, the sign of ${\cal Q}$ is irrelevant---see \eqref{Cl_TT2} and \eqref{f_ell}. However, care must be taken when considering the cross-angular power spectrum, where different redshift patches $\Delta{z}$ are cross correlated---as the sign of ${\cal Q}$ may vary in different $\Delta{z}$, and hence could affect the output of the prediction.

In Fig.~\ref{fig:1} we show the plot of the radial angular power spectrum of the magnification overdensity, with all the GR corrections taken into account, i.e.~for $\Delta^{\rm obs}_{_{\cal M}}$ \eqref{Delta:obs2}, and for the standard term containing only the weak lensing effect, i.e.~for $\Delta^{\rm std}_{_{\cal M}}$ \eqref{MagStd}---at the epoch $z_{_S}=0.1$. We see that at this epoch, the full (GR-corrected) power spectrum $C_\ell$ is greater in power than the standard (lensing) power spectrum $C^{\rm std}_\ell$, by a factor $C_\ell / C^{\rm std}_\ell \sim 10^3$. This difference is mainly owing to the Doppler effect in $C_\ell$; the Doppler term in $\Delta^{\rm obs}_{_{\cal M}}$ dominates at low $z$~\cite{Bonvin:2008ni, Bacon:2014uja}, which fluctuates on small $\ell \lesssim 100$. Our results are also in agreement with the work by \cite{Bonvin:2011bg} (see Fig.~$3$, top panel, by~\cite{Bonvin:2011bg}). Clearly, we see that the effect of GR corrections in the magnification power spectrum at the given epoch is about a thousand times in excess of the weak lensing effect---which may allow for the measurement of the GR effects. Thus the magnification power spectrum not only lends another avenue to study GR effects, but also offers a good possibility to measure GR effects at low $z$, on large scales. In contrast, the combined contribution of the GR effects in the observed galaxy power spectrum at low $z$ is largely subdominant---hence may be difficult to measure at low $z$. (Morover, for a single-tracer two- or three-dimensional galaxy power spectrum, all previously undetected GR corrections---i.e.~excluding weak lensing---are completely unobservable~\cite{Alonso:2015uua}.)

\begin{figure}\centering
\includegraphics[scale=0.4]{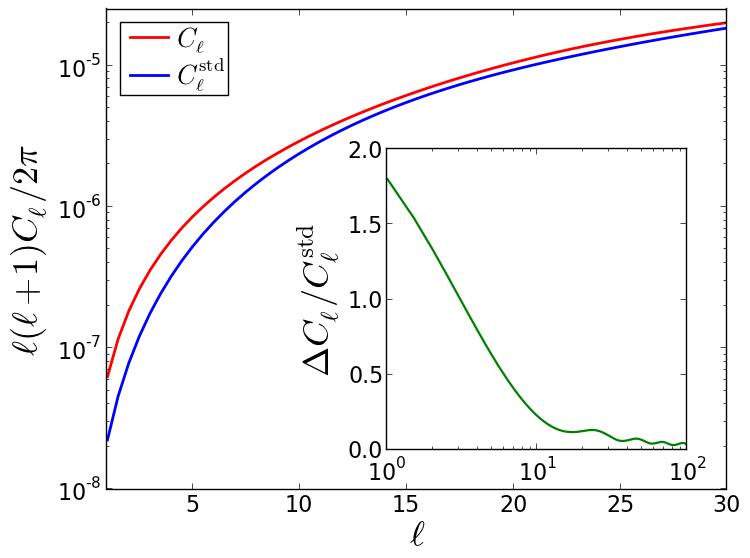}\\[-4.5mm] \includegraphics[scale=0.4]{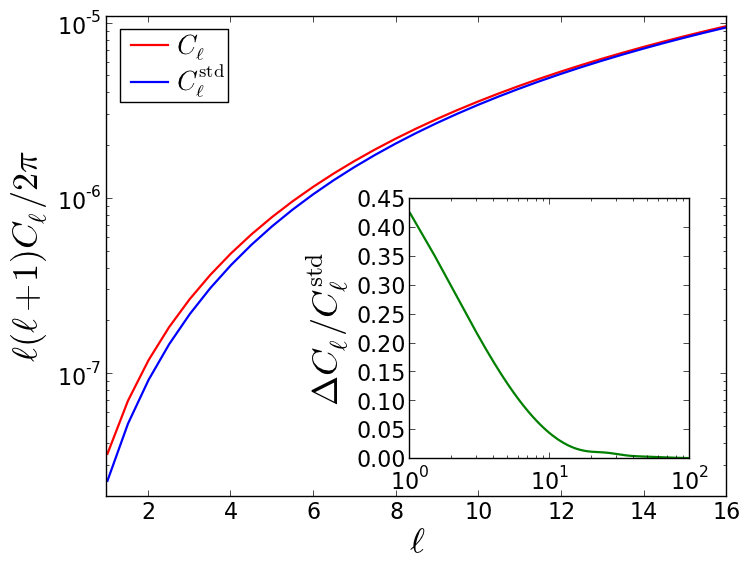} 
\caption{The magnification (radial) angular power spectrum with ${\cal Q}=1$, for noninteracting DE scenario: at $z_{_S}=1$ (top panel), and at $z_{_S}=3$ (bottom panel). Line styles are as in Fig.~\ref{fig:1}. The insets show the fractional changes in the angular power spectrum at the given $z$, where $\Delta C_\ell \equiv C_\ell - C^{\rm std}_\ell$.}\label{fig:2}
\end{figure}
    
Similarly, in Fig.~\ref{fig:2} we give the plot of the radial angular power spectrum of the magnification overdensity, at $z_{_S}=1$ (top panel), and at $z_{_S}=3$ (bottom panel). We see that at the given epochs, the amplitude of the weak lensing power spectrum $C^{\rm std}_\ell$ approaches that of the GR magnification power spectrum $C_\ell$. This implies that at $z \geq 1$, the weak lensing effect in the magnification angular power spectrum gradually becomes significant. We observe (Figs.~\ref{fig:1} and~\ref{fig:2}) that there is a consistent decrease in the amplitude of $C_\ell$ with increasing $z$; with the contribution of the GR effects (relative to the weak lensing effect) gradually falling, to ${\sim} 40\%$ at $z_{_S}=3$ (see inset), which is a significant amount nevertheless, as we enter an era of precision cosmology---e.g. BOSS is expected to measure the area distance $\tilde{D}_A$ with a precision of ${\sim}1.0\%$ at $z < 0.7$ and ${\sim}4.5\%$ at $z \approx 2.5$ (with higher $\%$ at $2 \lesssim z \lesssim 3.5$) \cite{Eisenstein:2011sa}, while the SKA is expected to be better (${\sim}0.3\%$ at $z\approx 1.3$) \cite{Yahya:2014yva}. (Note however that, in reality, detecting the actual effect of the GR corrections depends on the cosmic variance on the given scales, and the error bars achievable by the survey experiment; but for the purpose of this work, we leave out all exact experimental aspects.) In general, given the large relative contribution of the GR effects it implies that even at low $z$, by using the magnification power spectrum, GR effects can be suitably probed (and, in principle, measured)---contrary to the case of the galaxy power spectrum, which requires going to very high $z$ (and large magnification bias).

\section{The Power Spectrum with Interacting Dark Energy}\label{IDE:Case}% 
The dark sector, i.e.~DE and DM, does not interact with baryonic matter. In the standard cosmologies, i.e.~as considered in Sec. \ref{MagPk}, baryons, DM and DE interact only indirectly by gravitation (via the Poisson equation). However, DE may interact with DM non-gravitationally, via a reciprocal exchange of energy and momentum; thus, is called interacting DE (IDE) \cite{Duniya:2015ths, Duniya:2015nva, Gavela:2010tm, Salvatelli:2013wra, Costa:2013sva}. In this section we probe the magnification angular power spectrum for an IDE scenario---assuming (hereafter) a late-time universe dominated by DM and DE only.

\subsection{The IDE model}\label{IDE:mod}%
We assume that the energy density transfer $4$-vectors $Q^\mu_A$ ($A=m$, $x$, denoting DM and DE, respectively) are parallel to the DE $4$-velocity:
\bea\label{trans:Case}
Q^\mu_x \;=\; Q_x u^\mu_x \;=\; -Q^\mu_m,
\eea 
i.e.~there is zero momentum transfer in the DE rest frame; $Q_x$ is the DE (energy) density transfer rate, and $u^\mu_x$ is the DE 4-velocity. The momentum density transfer rates are
\beq\label{fm:fx}
f_x \;=\; \bar{Q}_x (V_x - V) \;=\; -f_m,
\eeq
where $V$ and $V_x$ are the total and the DE velocity potentials, respectively; the 4-velocities,
\bea\label{overDens:Vels}
u^\mu &=& a^{-1}\left(1 -\Phi,\, \partial^i V\right),\;  u^\mu_A \;=\; a^{-1}\left(1 -\Phi,\, \partial^i V_A\right),\quad\\ %\label{totalV}
V &=& \dfrac{1}{1+w}\dsum_A{\Omega_A\left(1+w_A\right)V_A},\;   w \;=\; \dsum_A{\Omega_A w_A},\nonumber
\eea
with $\Omega_A \equiv \bar{\rho}_A / \bar{\rho}$ being the density parameter, and $\bar{\rho}$ is the total background energy density. 

We specify the IDE model by choosing $Q_x$ \cite{Duniya:2015ths, Duniya:2015nva, Gavela:2010tm}:
\beq\label{Mod2:Q}
Q_x \;=\; \dfrac{1}{3}\xi \rho_x \Theta,\quad\quad \Theta \;=\; \nabla_\mu u^\mu,
\eeq
with the interaction parameter $\xi$ = constant, the DE (energy) density $\rho_x=\bar{\rho}_x + \delta\rho_x$ and, $\Theta$ the expansion rate:
\beq\label{nabla:u}
\Theta \;=\; 3a^{-1}\left[{\cal H} - \left(\Phi' + {\cal H}\Phi\right) +\dfrac{1}{3}\nabla^2 V\right].
\eeq
Note that, apart from \cite{Duniya:2015ths, Duniya:2015nva, Gavela:2010tm}, it is common in the literature to use an energy density transfer rate of the form $Q \propto a^{-1} {\cal H} \rho_x$, with the main motivation being that the background energy conservation equations are easily solved. However, the Hubble rate ${\cal H}$ is typically not perturbed---being a background parameter---which is thus a problem for the perturbed case of the given transfer rate. This problem is suitably resolved by \eqref{Mod2:Q}.
  
Equations \eqref{trans:Case}, \eqref{overDens:Vels}, \eqref{Mod2:Q} and \eqref{nabla:u} then lead to 
\begin{align*}
Q_x =&\; \bar{Q}_x \Big[1 +\delta_x -\Phi -\dfrac{1}{3{\cal H} }\left(3\Phi' - \nabla^2 V\right)\Big] = -Q_m, \\
Q^x_\mu =&\; a\bar{Q}_x \left[ -1-\delta_x + \dfrac{1}{3{\cal H} }\left(3\Phi' - \nabla^2 V\right), \, \partial_i V_x\right] =-Q^m_\mu,
\end{align*}
where $\bar{Q}_x = a^{-1}\xi{\cal H}\bar{\rho}_x =-\bar{Q}_m$ are the DE and the DM background energy density transfer rates, respectively, and $\delta_x \equiv \delta{\rho}_x/\bar{\rho}_x$ is the DE density contrast. Moreover, the range of $w_x$ is restricted by stability requirements \cite{Duniya:2015ths, Duniya:2015nva, Salvatelli:2013wra, Costa:2013sva}
\beq\label{stab}
w_x > -1~~\mbox{for}~\xi > 0; ~~w_x < -1~~\mbox{for}~\xi < 0.
\eeq
We set the evolution equations such that, \eqref{stab} corresponds to the energy transfer directions:
\beq\label{etd}
\mbox{DM $\to$ DE for}~\xi > 0; ~~~\mbox{DE $\to$ DM for}~\xi < 0.
\eeq
(See \cite{Duniya:2015ths, Duniya:2015nva} for the full IDE background and perturbation evolution equations.)

\subsection{The $C_\ell$'s with IDE}\label{IDE:Cls}% 
Here we probe the magnification (radial) angular power spectrum in a universe with IDE, for various values of the interaction parameter. The overall behaviour of the angular power spectra, i.e.~$C_\ell$ and $C^{\rm std}_\ell$, for the IDE scenario is similar to the standard DE scenario (Figs.~\ref{fig:1} and \ref{fig:2})---except that the power is suppressed. The chosen values of $\xi$ are such that DM transfers energy and momentum to DE---see~\eqref{stab} and~\eqref{etd}.

\begin{figure*}
\begin{tabular}{cc}
\includegraphics[scale=0.4]{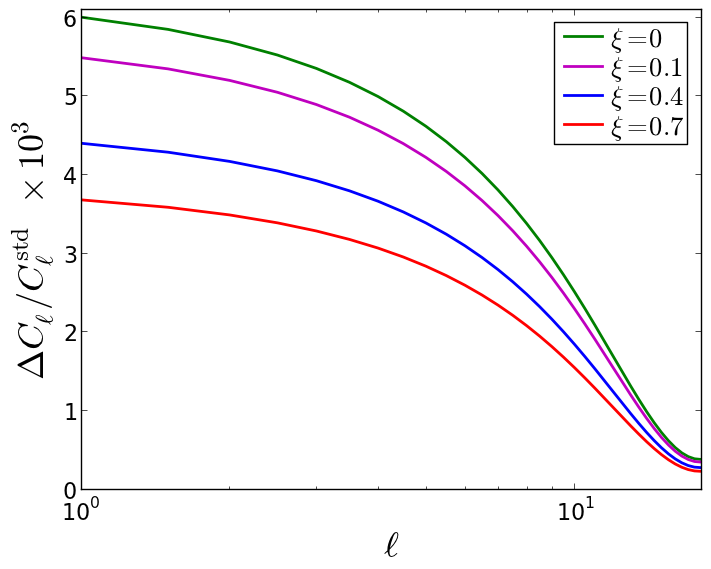} & \includegraphics[scale=0.4]{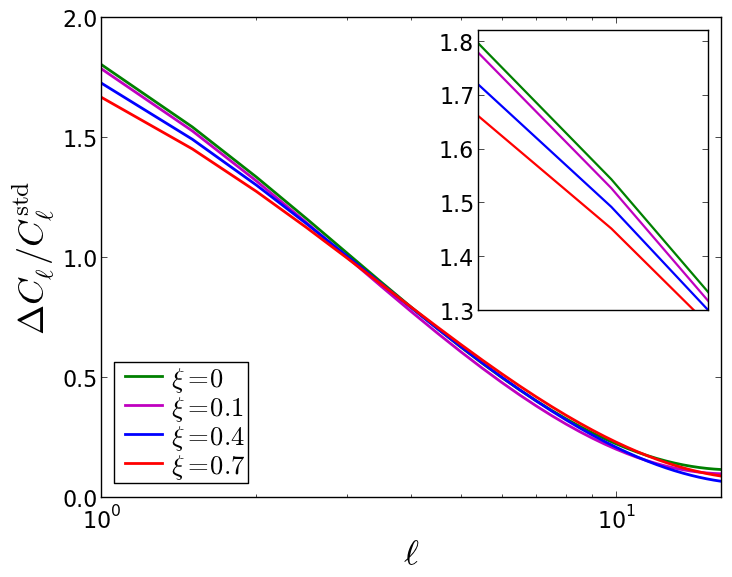} \\[-5mm]
\includegraphics[scale=0.4]{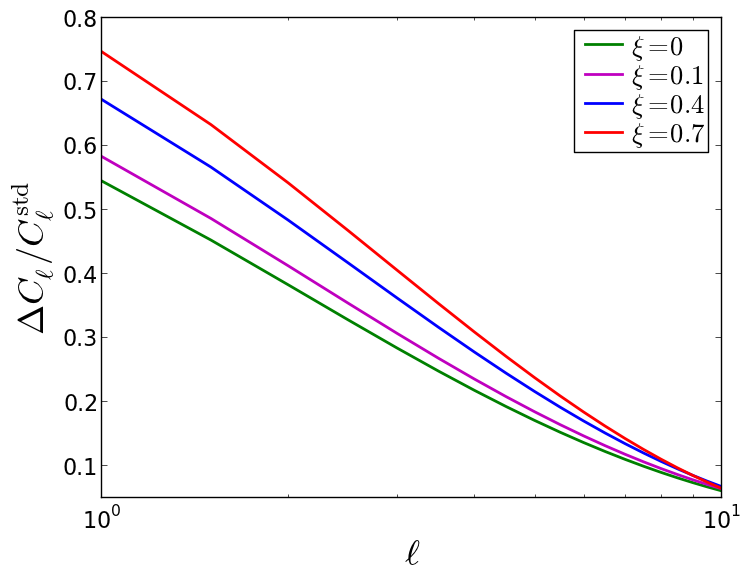} &\; \includegraphics[scale=0.4]{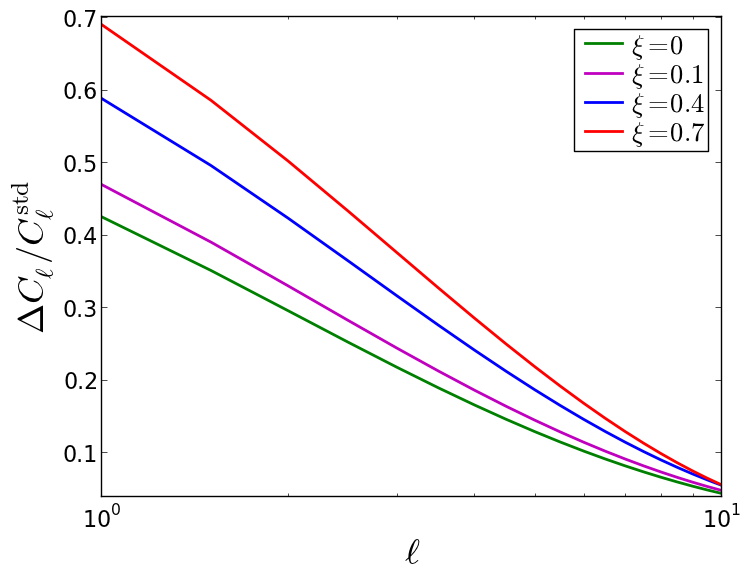} 
\end{tabular}
\caption{The fractional change---owing to GR effects---in the magnification angular power spectrum with IDE, for the following values of the interaction parameter $\xi =0,\, 0.1,\, 0.4,\, 0.7$: at $z_{_S}=0.1$ (top left), $z_{_S}=1$ (top right), $z_{_S}=2$ (bottom left) and $z_{_S}=3$ (bottom right). Notations are as in Fig.~\ref{fig:2}.}\label{fig:3}
\end{figure*}

In Fig.~\ref{fig:3}, we plot the fractional change $\Delta C_\ell / C^{\rm std}_\ell$ owing to the GR corrections in the magnification angular power spectrum, for the interaction parameter values $\xi=0,\, 0.1,\, 0.4,\, 0.7$: at $z_{_S}=0.1$ (top left panel), $z_{_S}=1$ (top right panel), $z_{_S}=2$ (bottom left panel) and $z_{_S}=3$ (bottom right panel). The ratios $\Delta C_\ell / C^{\rm std}_\ell$ for the various values of $\xi$ show the action or effect of the IDE on the GR effects in the magnification power spectrum. In both panels, we see that there is a consistent suppression of large-scale power (i.e.~on small $\ell$'s) in the magnification power spectrum---for larger values of $\xi \geq 0$ at epochs $z \leq 1$. This may be expected since DM loses energy (and momentum) to DE. Thus it implies that GR effects in the cosmic magnification at the given redshifts will diminish with increasing interaction strength, when DM transfers energy to DE. Note however that, at $z_{_S}=0.1$ the fractional contribution by the GR effects, i.e.~relative to the standard lensing effect, is still very high up to $\Delta C_\ell / C^{\rm std}_\ell \sim 10^3$ which is owing to the dominance of the Doppler effect at low $z$: the gravitational potential (which sources weak lensing) decays at low $z$---but grows as $z$ increases.

However, at $z_{_S}=1$ we see that the magnitude of the fractional change significantly falls to $\Delta C_\ell / C^{\rm std}_\ell \lesssim 2$, with a much smaller separation between successive lines (or fractions) on large scales; the amplitudes of the fractions at $z_{_S}=1$ fall by a factor of the order of $10^{-3}$, relative to the amplitudes at $z_{_S}=0.1$. This fall in amplitude is mainly due to the fact that as $z$ increases, the amplitude of the DM peculiar velocity (which sources the Doppler effect) decreases, via the lose of momentum on large scales. Thus on moving towards earlier epochs, the contribution of the Doppler effect---relative to the weak lensing effect---in the magnification power spectrum decreases. Moreover, the fact that we see relatively narrower separations between the fractions of the different values of $\xi \geq 0$ at $z_{_S}=1$, it implies that at this epoch the GR effects become less sensitive to the strength of the dark sector interaction. Thus trying to constrain the nature of IDE by GR effects (or vice versa), via the magnification power spectrum, at this epoch may not be suitable. Basically, the plots in the top panels (Fig.~\ref{fig:3}) show that IDE leads to the suppression of GR effects in the magnification power spectrum at $z\leq 1$---when DM loses energy and momentum to DE, the higher the rate of energy (and momentum) density transfer, the stronger the suppression.

\begin{figure*}
\begin{tabular}{cc}
\includegraphics[scale=0.4]{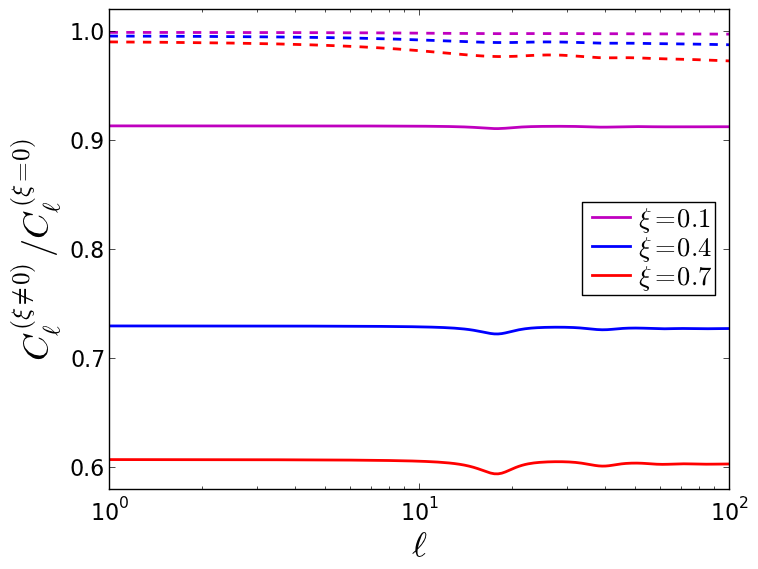} & \includegraphics[scale=0.4]{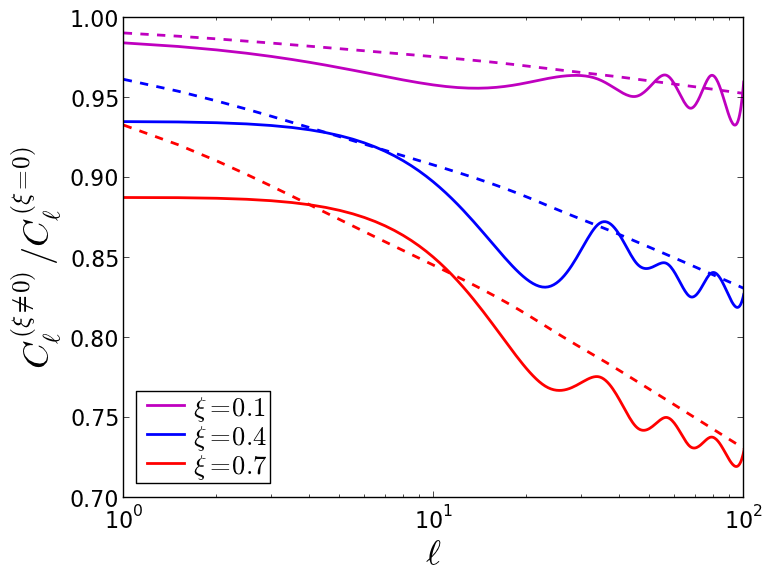} \\[-5mm]
\includegraphics[scale=0.4]{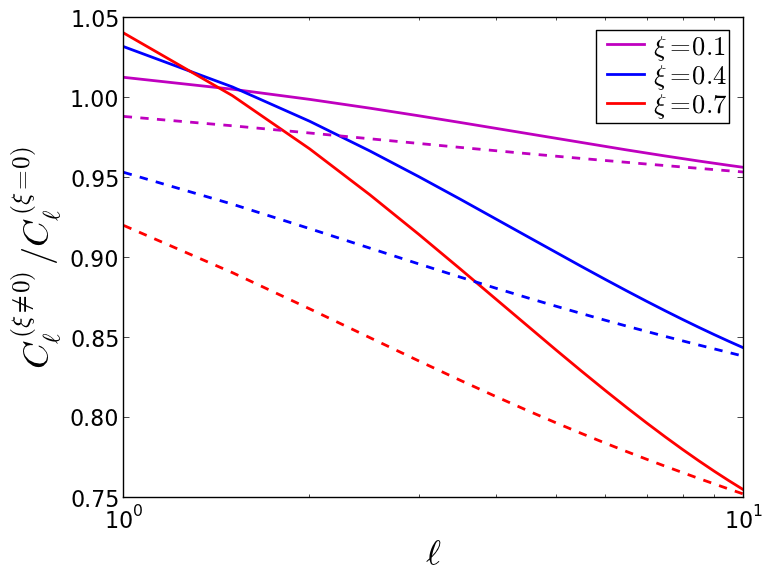} & \includegraphics[scale=0.4]{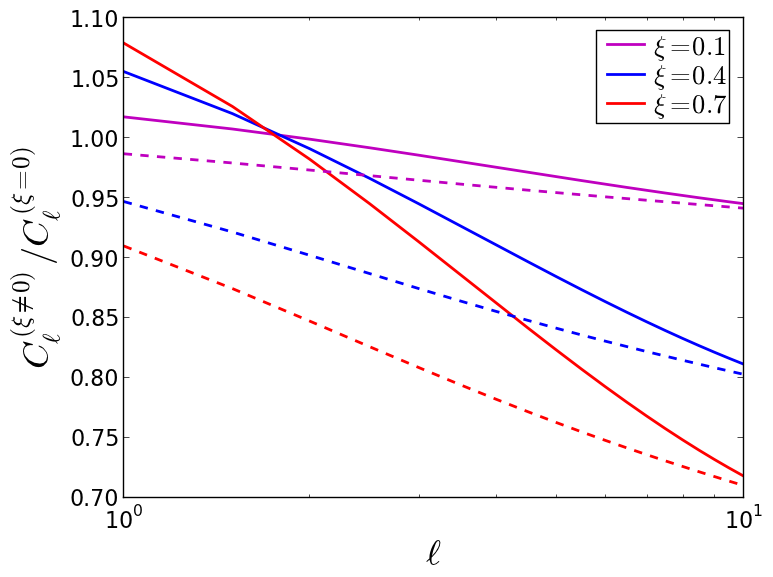} 
\end{tabular}
\caption{The ratios of the magnification (radial) angular power spectra: those with IDE ($\xi = 0.1,\, 0.4,\, 0.7$) relative to those with standard DE ($\xi=0$): at $z_{_S}=0.1$ (top left), $z_{_S}=1$ (top right), $z_{_S}=2$ (bottom left) and $z_{_S}=3$ (bottom right). The solid lines denote ratios of the full power spectrum $C_\ell$, while the dashed lines denote ratios of the standard (lensing) power spectrum $C^{\rm std}_\ell$.}\label{fig:4}
\end{figure*}

Moreover, in the bottom panels of Fig.~\ref{fig:3} (i.e.~at $z >1$), we observe that we have the converse behaviour of the plots in the top panels (i.e.~at $z \leq 1$): the fractional change $\Delta C_\ell / C^{\rm std}_\ell$ grows with increasing interaction strength, i.e.~the excess power induced by the GR effects increases as the rate of energy and momentum transfer between DM and DE increases. It is known that GR effects are typically stronger at high $z$, but with negative magnitude \cite{Duniya:2015nva}, i.e.~$\Delta^{\rm GR}_{_{\cal M}} < 0$ at high $z \gtrsim 1$. Moreover, given our metric choice, $\Delta^{\rm std}_{_{\cal M}} < 0$ for all $z$. Thus at high $z$ the correlation between $\Delta^{\rm GR}_{_{\cal M}}$ and $\Delta^{\rm std}_{_{\cal M}}$ leads to positive contribution in the magnification power spectrum, and hence a growing fraction $\Delta C_\ell / C^{\rm std}_\ell$ with increasing dark sector interaction strength. However, at high $z$ the IDE effects are weaker, since the effects of DE in general are weaker at earlier times; hence although GR effects become enhanced with increasing $\xi$, we see that the amplitude of each fraction (for a given value of $\xi \geq 0$) decreases as $z$ increases: compare the right and the left bottom panels in Fig.~\ref{fig:3}. At low $z$ we have $\Delta^{\rm GR}_{_{\cal M}} > 0$, so that its correlation with $\Delta^{\rm std}_{_{\cal M}}$ leads to negative contribution, thereby gradually reducing power in the magnification power spectrum for increasing $\xi \geq 0$, on the largest scales---which is the case in the top panels (Fig.~\ref{fig:3}). In essence, at $z \leq 1$ we have that IDE suppresses GR effects, while at $z >1$ an IDE supports the enhancement of GR effects in the magnification power spectrum---when DM loses energy and momentum to DE.

In Fig.~\ref{fig:4}, we show the plots of the ratios of the magnification angular power spectra, $C_\ell$ and $C^{\rm std}_\ell$: those with IDE (i.e.~$\xi \neq 0$) relative to those with standard DE (i.e.~$\xi = 0$); at the source epochs $z_{_S}=0.1$ (top left panel), $z_{_S}=1$ (top right panel), $z_{_S}=2$ (bottom left panel) and $z_{_S}=3$ (bottom right panel). These results show the IDE effects in the magnification angular power spectrum---with and without GR effects. The ratios of $C^{\rm std}_\ell$ (dashed lines) show the effect purely from the IDE; we see, in the four panels, that IDE leads to power suppression on all scales in the standard magnification power spectrum. There is a consistent suppression of power for increasing $\xi >0$, with the ratios gradually growing from small scales, tending to converge on the largest scales---such that the rate of convergence increases, on moving towards the present epoch. Moreover, the amplitude of the various ratios decreases very slowly as $z$ increases, supporting the fact that the IDE effect is weaker at higher $z$. However, on introducing the GR effects we see significant changes in the behaviour of the ratios, i.e.~the ratios of $C_\ell$ (solid lines)---which measure the IDE effect in the presence of GR effects. At $z_{_S}=0.1$ we see that, with GR effects, the ratios become well differentiated. This implies that GR effects cause the IDE effect to become more prominent, and sensitive on large scales. This will be crucial for constraints on IDE. On the other hand, the associated ratios of $C^{\rm std}_\ell$ show weak sensitivity to the IDE effect, having relatively negligible separations. This implies that at near epochs $z \ll 1$, the standard (lensing) magnification power spectrum will not be suitable for constraints on IDE on very large scales.

On going from $z_{_S}=0.1$ through to $z_{_S}=3$ (i.e.~top left to bottom right panels) we see how the GR corrections influence the IDE effect in the magnification angular power spectrum, on the largest scales. For a given value of the interaction parameter $\xi>0$, at late epochs $z \lesssim 1$ the IDE effect is reduced (and lower) when GR corrections are included; while at early epochs $z>1$ the IDE effect becomes enlarged (and higher) when GR corrections are included---however with the IDE effect becoming well differentiated, and prominent in all cases. Thus this implies that if GR corrections are not taken into account in the analysis, the IDE effect will not be properly illuminated (and/or incorporated), which may lead to an incorrect estimate of the large-scale imprint of IDE in the cosmic magnification. Including the GR corrections may also present the possibility of discriminating the IDE effect from any other (possible) large-scale effects in the cosmic magnification. Thus by neglecting GR corrections, the true potential of the cosmic magnification as a cosmological probe may be severely reduced (or forfeited).

\section{Conclusion}\label{Conc}
We have investigated GR effects in the observed cosmic magnification power spectrum. After re-deriving the known GR magnification overdensity, we discussed the GR effects in noninteracting DE scenario---where we compared the full GR-corrected magnification radial angular power spectrum with the (standard) lensing magnification angular power spectrum. In a similar manner, we probed the magnification angular power spectrum with IDE. Furthermore, we compared the angular power spectra of the IDE scenario with those of the noninteracting DE scenario, throughout keeping the DE physical sound speed $c_{sx}=1$, and a magnification bias ${\cal Q} = 1$. (Note however that given the purpose of this work, the value and/or form of ${\cal Q}$ is irrelevant---as its effect is cancelled out in the power spectrum ratios.)

We found that for the standard DE scenario, while the weak lensing effect in the magnification power spectrum grows as redshift $z$ increases, the total contribution by the GR effects---i.e.~relative to the sole weak lensing effect---falls gradually, to about $40\%$ at $z=3$ on very large scales (which is a significant amount, especially as we enter the era of precision cosmology). Moreover, we found that the magnification power spectrum can be suitably used to probe (and in principle, measure) GR effects at low $z$---contrary to the case of the galaxy power spectrum, which requires going to very high $z$. In essence, the cosmic magnification offers a better means of elaborating the effects of GR corrections (and DE, in general).

We also found that IDE suppresses the GR effects in the magnification angular power spectrum at epochs $z \leq 1$, when DM loses energy (and momentum) to DE: the higher the rate of energy transfer, the stronger the suppression. Whereas at $z > 1$, the contribution of GR effects become enhanced with increasing interaction strength. This is because at high $z$, the correlation between the GR term and the weak lensing term has a positive contribution in the magnification power spectrum---which grows with increasing $z$; while at low $z$, this term gives a negative contribution (consequently reducing the power amplitude). 

The IDE effect generally showed a strong sensitivity to the GR corrections in the magnification power spectrum, on large scales---which will be crucial for constraints on IDE, particularly at low $z$. Moreover, the results showed that the IDE effect becomes more elaborate, and prominent when GR corrections are included; thus if GR corrections are omitted in the analysis, this may lead to an incorrect estimate of the large-scale imprint of IDE in the cosmic magnification. Including the GR corrections can enhance the true potential of the cosmic magnification as a cosmological probe.

%%%%%%%%%%%%%%%%%%%%%%%%%%%%%%%%%%%%%%%%%%%%%%%%%%%%%%%%%%%%%%%%%%%%%%%%%%%%%%%%%%%%%%%%%
\[\]{\bf Acknowledgements:} Thanks to Roy Maartens and Chris Clarkson for meaningful discussions. This work was carried out with financial support from (1) the South African Square Kilometre Array Project and the South African National Research Foundation, (2) the government of Canada's International Development Research Centre (IDRC), and within the framework of the AIMS Research for Africa Project. 
%%%%%%%%%%%%%%%%%%%%%%%%%%%%%%%%%%%%%%%%%%%%%%%%%%%%%%%%%%%%%%%%%%%%%%%%%%%%%%%%%%%%%%%%%

%=========================================================================================
\appendix 
%=========================================================================================

\section{Derivation of the Magnification Overdensity}\label{MagDistortn}
All derivations in this appendix---which give some of the details of Sec.~\ref{sec:DeltaMag}---are taken from the more rigorous work by \cite{Duniya:2015ths} (and references therein); assuming flat space throughout.

\subsection{The metric}\label{subsec:pertgeodesic}
The metric is often expressed in the form of a quadratic differential, given in terms of the geometric metric tensor $g_{\mu\nu}$, in real coordinates $x^\mu$ by
\beq\label{Metric}
ds^2 \;=\; g_{\mu\nu}dx^{\mu}dx^{\nu} .
\eeq 
In a perturbed Friedmann-Robertson-Walker universe, the metric tensor may be decomposed as follows:
\beq\label{MetricTens}
g_{\mu\nu} \;=\; \bar{g}_{\mu\nu} + \delta g_{\mu\nu},
\eeq
where $\bar{g}_{\mu\nu} = \bar{g}_{\mu\nu}(\bar{\eta})$ is the background term, $\delta g_{\mu\nu} = \delta g_{\mu\nu}(\eta,x^i)$ is the perturbation, with $\eta=\bar{\eta}+\delta\eta$ being the conformal time, and
\beq\label{BKD:MetricTens}
\bar{g}_{00} = -a^2,\quad \bar{g}_{i0} = \vec{0} = \bar{g}_{0j},\quad \bar{g}_{ij} = a^2\delta_{ij},
\eeq
where we consider (henceforth) only linear perturbations; $a$ is the scale factor. The perturbation $\delta g_{\mu\nu}$ may be parametrized by scalar fields, i.e.~if ${x}^{i}$ denotes the space $3$-vector, then we can express the perturbation of the metric tenor by the scalar quantities $\phi =\phi(\eta ,{x}^{i})$, $B =B(\eta ,{x}^{i})$, $D =D(\eta ,{x}^{i})$ and $E =E(\eta ,{x}^{i})$, given by
\beq\nonumber%\label{PertMetricComps}
\delta{g}_{00} = -2a^{2}{\phi},\; \delta{g}_{i0} = a^2B_{i},\; \delta{g}_{ij} = -2a^2\left(D\delta_{ij}-E_{ij}\right),
\eeq
where $B_{i} = B_{|i}$ and $E_{ij} = E_{|ij} -\frac{1}{3}\delta_{ij}\nabla^2E$ is a traceless transverse tensor---i.e.~$E^i\/_i=0$, such that it has no contribution to the term, $D\delta^i\/_i$, in the diagonal plane. We denote $X_{|i} \equiv \nabla_iX$, and $X_{|ij} \equiv \nabla_i\nabla_jX$ for a scalar $X$.

Henceforth we adopt the conformal transformation:
\bea\label{Conf:metric}\nonumber
ds^2 \to d\tilde{s}^2 &=& a^2 ds^2,\\ \label{Conf:metric2}
&=& a(\eta)^2 \left\lbrace -\left(1+2\phi\right) d{\eta}^2 + 2B_{|i}\, d{\eta} dx^i \right.\nn
&& + \left. \left[ (1 - 2\psi)\delta_{ij} +2E_{|ij}\right] dx^i dx^j \right\rbrace ,
\eea 
where $\psi \equiv D +\frac{1}{3}\nabla^2E$, and we have assumed flat space. Note that all the given scalar amplitudes of the metric~\eqref{Conf:metric2} perturbations are coordinate-dependent. Thus, \eqref{Conf:metric2} implies that the respective metric tensors are %related by
\bea\label{metrictens9}
\tilde{g}_{\mu\nu} &=& a^2 \left(\bar{g}_{\mu\nu} + \delta{g}_{\mu\nu}\right),\quad \bar{g}_{00} =  -1,\nn
\bar{g}_{i0} &=&\vec{0}=\bar{g}_{0j},\quad \bar{g}_{ij} = \delta_{ij}.
\eea
where an overbar denotes background component. For a geodesic $\tilde{x}^\nu(\lambda)$ in the metric $d\tilde{s}$, the associated tangent vectors are given by
\beq\label{n:Conf}
\tilde{n}^\mu \;=\; a^{-2}n^\mu \;=\; a^{-2}(1+\delta{n}^0,\, \bar{n}^i +\delta{n}^i),
\eeq
where $n^\mu = dx^\mu / d\lambda$ and $\lambda$ is the affine parameter. Henceforth, we assume photon (or null) geodesics: hence $\bar{n}^\mu \bar{n}_\mu = 0$, with $\bar{n}^0\bar{n}_0=-1$ (where $\bar{n}^0=1$) and $\bar{n}^i \bar{n}_i=1$. The $4$-velocities of a particle moving in $d\tilde{s}$, are given by
\bea\label{vels}
\tilde{u}^\mu &=& a^{-1} u^\mu \;=\; a^{-1}\left(1 -\phi,~v_{|i}\right),\nn 
\tilde{u}_\mu &=& a\, u_\mu \;=\; a\left(-1 -\phi,~v_{|i}+B_{|i}\right),
\eea
where $v_{|i} = \partial_i v$, and $v$ is the velocity (scalar) potential.

\subsection{Gauge-invariant potentials}\label{subsec:GIP}
We have the well-known Bardeen potentials $\Phi$ and $\Psi$, and the gauge-invariant velocity potential $V$, given by
\bea\label{InvPhi}
\Phi &\;\equiv\;& \phi -{\cal H} \sigma -\sigma',\\ \label{InvPsi}
\Psi &\;\equiv\;& D +\dfrac{1}{3} \nabla^2E + {\cal H} \sigma, \\ \label{InvVel}
V &\;\equiv\;& v + E',
\eea
where $\sigma = -B+E'$. These correspond to the potentials in conformal Newtonian gauge.

\subsection{The position $4$-vector}
The position 4-vector ${x}^\mu$ of a photon moving in the direction $\bn$, from a given source $S$ to an observer $O$, is
\beq\label{posVector}
x^\mu(\bar{\eta}_{_S}) \;=\; -\left(\bar{\eta}_{_O} - \bar{\eta}_{_S}\right)\bar{n}^\mu - \int^{\bar{r}_{S}}_0{d\bar{r} \left( \delta{n}^\mu - \bar{n}^\mu\delta{n}^0\right)},
\eeq
where $\bar{r}_{_S} = \bar{r}(\bar{\eta}_{_S})$ with $\bar{r}(\bar{\eta}_{_O}) = 0$, and to lowest order along the photon geodesic
\beq\label{eq:dr}
d\bar{\eta} \;=\; -d\bar{r} \;=\; d\lambda. 
\eeq
Thus we have the position deviation $4$-vector, given by
\bea\label{DeviatnVec}
\delta{x}^i(\bar{\eta}_{_S}) &=& \dfrac{1}{2} \int^{\bar{r}_S}_0{ d\bar{r}  (\bar{r}_{_S} - \bar{r}) \left(\bar{g}^{ij}\partial_j\delta{g}_{\alpha\beta} + \delta{g}'_{\alpha\beta}\bar{n}^i \right) \bar{n}^\alpha \bar{n}^\beta }\nn 
&& + \int^{\bar{r}_S}_0{ d\bar{r} \left( \bar{g}^{ij}\delta{g}_{j\beta} + \delta{g}_{0\beta}\bar{n}^i\right) \bar{n}^\beta } ,
\eea
where we used the following identities:
\bea\label{pert_n0}
\delta{n}^0 &\;=\;& \delta{g}_{0\beta}\, \bar{n}^\beta - \dfrac{1}{2} \int^0_{\bar{r}_S}{ d\bar{r}\, \delta{g}'_{\alpha\beta}\, \bar{n}^\alpha\bar{n}^\beta } , \\ \label{pert_ni}
\delta{n}^i &\;=\;& - \bar{g}^{ij}\delta{g}_{j\beta}\, \bar{n}^\beta + \dfrac{1}{2} \bar{g}^{ij} \int^0_{\bar{r}_S}{ d\bar{r}\, \partial_j \delta{g}_{\alpha\beta}\, \bar{n}^\alpha\bar{n}^\beta } ,
\eea
where $\delta{n}^\mu \equiv \delta{n}^\mu |^{^S}_{_O}$. See \cite{Duniya:2015ths, Bonvin:2011bg} for further details regarding the calculations in this subsection.

\subsection{The transverse area}
Here we compute the area density transverse to the photon geodesic. From \eqref{A:defn}, we have the only nonvanishing terms to yield (and the indices $i$, $j$, $k$ and $l$ denote spatial components)
\bwt
\bea%\label{DefnA1}
\mathpzc{A} &=& a^{-2}\sqrt{-\tilde{g}} \left[1 + \dfrac{\delta{u}^0}{\bar u^0} + \dfrac{\delta{\ell}^l}{\bar{\ell}^l}\right] \epsilon_{ijk}\,\bar{\ell}^i\, \dfrac{\partial\tilde{x}^j}{\partial\theta_{_S}} \dfrac{\partial\tilde{x}^k}{\partial\vartheta_{_S}} \left| \left. \dfrac{\partial(\theta_{_S},\vartheta_{_S})}{\partial(\theta_{_O},\vartheta_{_O})} \right| \right. ,\nn %\label{DefnA2}
&=& a^2 r^2\sin{\tilde \theta_{_S}} \left[1 -3D -\phi + \bar{n}^l B_{|l} - \dfrac{1}{2}\delta{g}_{\mu\nu}\bar{n}^\alpha\bar{n}^\beta\right] \left| \left. \dfrac{\partial(\tilde{\theta}_{_S},\tilde{\vartheta}_{_S})}{\partial(\theta_{_O},\vartheta_{_O})} \right| \right., \nn \label{DefnA3}
&=& a^2\bar{r}^2\sin{\theta_{_O}}\left[1 -3D -\phi + \bar{n}^i B_{|i} - \dfrac{1}{2}\delta{g}_{\mu\nu}\bar{n}^\alpha\bar{n}^\beta +2\dfrac{\delta{r}}{\bar{r}} + \left(\cot{\theta_{_O}} +\partial_\theta\right)\delta{\theta} +\partial_\vartheta\delta{\vartheta}\right],\quad
\eea
\ewt 
where $\sqrt{-\tilde{g}} =a^4(1+\phi-3D)$, with $\bar{\tilde u}^\mu = a^{-1}\delta^\mu\/_0$ and $\bar{\tilde u}_\mu = -a\delta^0\/_\mu$ as given by \eqref{vels}. The determinant of the transformation matrix becomes $|\partial(\theta_{_S},\vartheta_{_S}) / \partial(\theta_{_O},\vartheta_{_O})| = 1 + \partial_\theta\delta{\theta} +\partial_\vartheta\delta{\vartheta}$, with $\partial_\theta \equiv \partial / \partial{\theta_{_O}}$; for $|\delta\theta| \ll 1$, we have $\sin\theta_{_S} = (1 + \delta{\theta}\cot{\theta_{_O}}) \sin{\theta_{_O}}$. Also, \eqref{ell:defn} becomes
\beq
\tilde{\ell}^\nu = a^{-1} \left(u^\nu + \dfrac{n^\nu}{n^\alpha u_\alpha} \right) =  a^{-1}\ell^\nu .
\eeq
Given \eqref{n:Conf} and \eqref{vels} we have $\bar{\ell}^0 = 0$ and $\bar{\tilde \ell}^i = a^{-1}\bar{\ell}^i = -a^{-1}\bar{n}^i / \bar{n}^0$. Then we obtain
\bea
\dfrac{\delta{\ell}^i}{\bar{\ell}^i} &=& -\bar{n}_i v^{|i} + \bar{n}_i\delta{n}^i - \delta{n}^0 + \delta{u}^0 - \bar{\ell}^i\delta{u}_i, \nn
&=& \bar{n}^i B_{|i} - \phi - \dfrac{1}{2} \delta{g}_{\alpha\beta} \bar{n}^\alpha\bar{n}^\beta,
\eea
where in the first line we used the identity $\bar{n}_i = 1/\bar{n}^i$, and the second line comes by combining \eqref{pert_n0} and \eqref{pert_ni} and integrating once. 

To compute the various terms of \eqref{DefnA3}, we need to relate polar coordinates $\check{x}^\mu$ to Cartesian coordinates $x^\mu$. The deviation $4$-vectors are related (to first order) by
\beq\label{polCartTrans}
\delta\check{x}^\mu \;=\; \dfrac{\partial\check{x}^\mu}{\partial{x}^\nu} \delta{x}^\nu \;=\; \delta^\mu\/_\nu\, \delta{x}^\nu.
\eeq
An infinitesimal deviation in the position of a photon is given by the $3$-vector
\beq\label{polarDevVect}
\boldsymbol{\delta \check{x}} \;=\; \delta{r}\, {\pmb e}_r + \bar{r}\, \delta{\theta}\, {\pmb e}_{\theta} + \bar{r}\sin{\theta} \, \delta{\vartheta}\, {\pmb e}_{\vartheta}, 
\eeq
where ${\pmb e}_r$, ${\pmb e}_{\theta}$ and ${\pmb e}_{\vartheta}$ are the orthonormal unit vectors of the polar coordinates $\check{x}^\mu$, with
\beq
{\pmb e}_{\theta} = \partial_{\theta} {\pmb e}_r,\quad {\pmb e}_{\vartheta} \sin{\theta} = \partial_{\vartheta} {\pmb e}_r,\quad {\pmb e}_r = -\bn,
\eeq
where ${\pmb e}_r\cdot {\pmb e}_r = {\pmb e}_{\theta}\cdot {\pmb e}_{\theta} = {\pmb e}_{\vartheta}\cdot{\pmb e}_{\vartheta} = 1$, and ${\pmb e}_r \cdot {\pmb e}_{\theta} = {\pmb e}_{\theta} \cdot {\pmb e}_{\vartheta} = {\pmb e}_{\vartheta} \cdot {\pmb e}_{r} = 0$. From~\eqref{polarDevVect}, we get
\beq\label{VectCompts}
\delta{r} = {\pmb e}_{r}\cdot\boldsymbol{\delta\check{x}}, \; \bar{r}\,\delta{\theta} = {\pmb e}_{\theta}\cdot \boldsymbol{\delta\check{x}},\; \bar{r}\sin{\theta}\,\delta{\vartheta} = {\pmb e}_{\vartheta}\cdot \boldsymbol{\delta\check{x}}.
\eeq
Moreover, the components of the Laplacian in spherical coordinates are given by
\bea\label{nablaCompts}
\partial{r} = -\bar{n}^i \partial_i,\quad \dfrac{1}{\bar{r}}\partial_{\theta} = e^i_{\theta} \partial_i,\quad \dfrac{1}{\bar{r}\sin{\theta}}\partial_{\vartheta} = e^i_{\vartheta} \partial_i  ,
\eea
Thus given~\eqref{DeviatnVec} and~\eqref{polCartTrans}, we get that (but see \cite{Duniya:2015ths, Bonvin:2011bg} for details) 
\bea\label{delta_r2}
\delta{r} &=& -\dfrac{1}{2} \int^{\bar{r}_S}_{0}{ d\bar{r}\, \delta{g}_{\alpha\beta} \, \bar{n}^\alpha \bar{n}^\beta }, \nn
&=& \int^{\bar{r}_S}_0{d\bar{r} \left(\Phi + \Psi\right)} + \left[B + \left(\dfrac{dE}{d\lambda} - 2E'\right) \right]^S_O ,
\eea
i.e.~$\delta{r} = -\bar{n}_i\, \delta{x}^i$, where we have integrated by parts once and applied the stationary condition on surface terms, which then vanish. Similarly, we have
\bea\label{delta_theta}
\bar{r}_{_S}\,\delta{\theta} &=& e_{\theta i}\,\delta{x}^i, \nn 
 &=& -\dfrac{1}{2} \int^{\bar{r}_S}_0{ d\bar{r} (\bar{r} - \bar{r}_{_S}) e^j_{\theta}\,\partial_j \left(\delta{g}_{\alpha\beta} \right)\bar{n}^\alpha \bar{n}^\beta }\nn
 && +\int^{\bar{r}_S}_0{ d\bar{r}\, \delta{g}_{j\beta} \, e^j_{\theta}\, \bar{n}^\beta } , \\\label{delta_vartheta}
\bar{r}_{_S}\,\sin{\theta}\,\delta{\vartheta} &=& e_{\vartheta i}\,\delta{x}^i, \nn
&=& -\dfrac{1}{2} \int^{\bar{r}_S}_0{ d\bar{r} (\bar{r} - \bar{r}_{_S}) e^j_{\vartheta}\,\partial_j \left(\delta{g}_{\alpha\beta} \right) \bar{n}^\alpha \bar{n}^\beta } \nn
 && +\int^{\bar{r}_S}_0{ d\bar{r}\, \delta{g}_{j\beta} \, e^j_{\vartheta}\, \bar{n}^\beta } .
\eea
From \eqref{eq:dr} and~\eqref{delta_r2}, we have
\beq
\dfrac{d\delta{r}}{d\lambda} = \dfrac{1}{2}\delta{g}_{\alpha\beta}\, \bar{n}^\alpha \bar{n}^\beta = -\left(\Phi + \Psi\right) + \dfrac{dB}{d\lambda} + \left(\dfrac{d^2E}{d\lambda^2} -2\dfrac{dE'}{d\lambda}\right),
\eeq
where we used~\eqref{InvPhi} and \eqref{InvPsi} and that 
\beq\label{dXdlam}
\dfrac{dX}{d\lambda} \;=\; X' + \bar{n}^i\partial_i{X} \;=\; X' - \partial_r{X},
\eeq
where $X$ is a scalar, and $\partial_r \equiv \partial / \partial{r}$ is the partial derivative with respect to $r$. After some lengthy, but straightforward calculations (see \cite{Duniya:2015ths, Bonvin:2011bg}), we obtain 
\bea\label{delOm}
\left(\cot{\theta} + \partial_{\theta}\right) \delta{\theta} + \partial_{\vartheta}\delta{\vartheta} &=& \int^{\bar{r}_S}_0{ d\bar{r} \left(\bar{r}_{_S} - \bar{r}\right) \dfrac{\bar{r}}{\bar{r}_{_S}} \nabla^2_\perp \left(\Phi + \Psi\right)} \nn
&& -\left[\nabla^2_\perp E \right]^S_O,
\eea
where $\nabla^2_{\perp} \equiv\; \nabla^2 - \partial^2_r - 2\bar{r}^{-1} \partial_r$ is the ``screen-space'' Laplacian---i.e.~in the plane of the source (perpendicular to the line of sight); and we have used that fact that
\bea\label{dtheta}\nonumber
e^j_{\theta} \partial_j \left(\delta{g}_{\alpha\beta} \right) \, \bar{n}^\alpha \bar{n}^\beta &=& \dfrac{1}{\bar{r}}\, \left[ \partial_{\theta} \left(\delta{g}_{\alpha\beta} \, \bar{n}^\alpha \bar{n}^\beta \right) + 2\, \delta{g}_{\alpha j} \,\bar{n}^\alpha e^j_{\theta} \right], \\\label{dvartheta}
e^j_{\vartheta} \partial_j \left(\delta{g}_{\alpha\beta} \right) \, \bar{n}^\alpha \bar{n}^\beta &=& \dfrac{1}{\bar{r}\,\sin{\theta}}\, \Big[ \partial_{\vartheta} \left(\delta{g}_{\alpha\beta} \, \bar{n}^\alpha \bar{n}^\beta \right) \nn 
&&\hspace{1.5cm} +\; 2\, \delta{g}_{\alpha j} \,\bar{n}^\alpha e^j_{\vartheta} \sin{\theta} \Big]. \nonumber
\eea
Furthermore, we used the following terms, i.e.~given \eqref{nablaCompts} and \eqref{dXdlam},
\bea\label{eTheta}
\delta{g}_{\alpha j} \,\bar{n}^\alpha e^j_{\theta} &=& \dfrac{\partial_{\theta} B}{\bar{r}} + 2\bar{n}^ie^j_{\theta} E_{|ij}, \nn
&=& \dfrac{\partial_{\theta}B}{\bar{r}} + \dfrac{2}{\bar{r}} \partial_{\theta}\left[\dfrac{dE}{d\lambda} - E'\right] .
\eea
Then in a similar manner, we obtain that
\beq\label{eVarTheta}
\delta{g}_{\alpha j} \,\bar{n}^\alpha e^j_{\vartheta} \;=\; \dfrac{\partial_{\vartheta}B}{\bar{r}\sin{\theta}} + \dfrac{2}{\bar{r}\sin{\theta}} \partial_{\vartheta}\left[\dfrac{dE}{d\lambda} - E'\right].
\eeq
Moreover we have that
\beq\label{nabPerpE}
\nabla^2_{\perp}E \;=\; \nabla^{2}E - \left(\dfrac{d^2E}{d\lambda^2} -2\dfrac{dE'}{d\lambda} + E''\right) + \dfrac{2}{\bar{r}} \left(\dfrac{dE}{d\lambda} -E'\right).
\eeq
The perturbation in the redshift of the propagating photon is given by~\cite{Duniya:2015ths, Bonvin:2011bg} 
\beq\label{delz}
\dfrac{\delta{z}}{1+\bar{z}} \;=\; \left[\Phi + \Psi + \bn\cdot{\bf v} -\psi \right]^0_{z_S}  -\int^0_{r_S}{d\bar{r} \left(\Phi' + \Psi'\right) } .
\eeq
By using \eqref{Conf:metric2}, \eqref{metrictens9}, \eqref{InvPsi} and \eqref{dXdlam}, we get % \eqref{DefnA3}, we have that
\bea\label{delpsi2}
\delta_\phi &\equiv & -3D -\phi + \bar{n}^i B_{|i} - \dfrac{1}{2}\delta{g}_{\mu\nu}\bar{n}^\alpha\bar{n}^\beta,\nn
&=& -2\Psi -E +2{\cal H}\sigma - E'' - \left(\dfrac{d^2E}{d\lambda^2} -2\dfrac{dE'}{d\lambda}\right) .\quad
\eea
Then given \eqref{DefnA3}, \eqref{delta_r2}, \eqref{delOm}, \eqref{nabPerpE} and \eqref{delpsi2}
\bea\label{matA1}
\mathpzc{A} &=& a^2\bar{r}^2\sin{\theta_{_O}} \left[1 -2\Psi +\dfrac{2}{\bar{r}_{_S}} \int^{\bar{r}_S}_0{ d\bar{r} \left(\Phi +\Psi\right) } \right.\nn
&+& \left. \int^{\bar{r}_S}_0{ d\bar{r} \left(\bar{r} -\bar{r}_{_S}\right) \dfrac{\bar{r}}{\bar{r}_{_S}} \nabla^2_\perp \left(\Phi +\Psi\right) } + 2{\cal H}\left(1 - \dfrac{1}{\bar{r}_{_S}{\cal H}} \right) \sigma \right] .\nn
\eea
By taking a gauge transformation \eqref{perp_dA}, we get the redshift-space perturbation
\bea\label{dmatA1}
\tilde{\delta}_{\mathpzc{A}} &=& \dfrac{\mathpzc{A} - \mathpzc{\bar A}}{\mathpzc{\bar A}} ~-~ \dfrac{d\ln{\mathpzc{\bar A}}}{d\bar{z}} \delta{z}, \nn 
 &=& -2\Psi +\dfrac{2}{\bar{r}_{_S}} \int^{\bar{r}_S}_0{ d\bar{r} \left(\Phi +\Psi\right) } \nn
 && + \int^{\bar{r}_S}_0{ d\bar{r} \left(\bar{r} -\bar{r}_{_S}\right) \dfrac{\bar{r}}{\bar{r}_{_S}} \nabla^2_\perp \left(\Phi +\Psi\right) } \nn
&& +\; 2\left(1 - \dfrac{1}{\bar{r}_{_S}{\cal H}} \right) \left[{\cal H}\sigma + a\,\delta{z}\right],
\eea
where $\delta_{\mathpzc{A}} \equiv \delta\mathpzc{A}/\mathpzc{\bar A} = (\mathpzc{A} -\mathpzc{\bar A}) /\mathpzc{\bar A}$, with $\mathpzc{\bar A} \equiv a^2\bar{r}^2\sin{\theta_{_O}}$;
\beq
\dfrac{d\ln{\mathpzc{\bar A}}}{d\bar{z}} \;=\; 2a\left(\dfrac{1}{\bar{r}_{_S}{\cal H}} - 1\right) .
\eeq
Thus given \eqref{delz}, \eqref{dmatA1} and $\mu^{-1} = 1 + \tilde{\delta}_{\mathpzc{A}}$, we get
\bea\label{dmag1}
\mu^{-1}  &\;=\;& 1 -2\Psi +\dfrac{2}{\bar{r}_{_S}} \int^{\bar{r}_S}_0{ d\bar{r} \left(\Phi +\Psi\right) } \nn
&& -\; 2\left(1 - \dfrac{1}{\bar{r}_{_S}{\cal H}} \right) \left[\Phi +V_\parallel - \int^{\bar{r}_{_S}}_0{ d\bar{r} \left(\Phi' + \Psi'\right) }\right] \nn
&& + \int^{\bar{r}_S}_0{ d\bar{r} \left(\bar{r} -\bar{r}_{_S}\right) \dfrac{\bar{r}}{\bar{r}_{_S}} \nabla^2_\perp \left(\Phi +\Psi\right) } .
\eea
%%%%%%%%%%%%%%%%%%%%%%%%%%%%%%%%%%%%%%%%%%%%%%%%%%%%%%%%%%%%%%%%%%%%%%%%%%%%%%%%%%%%%%%%%%

%\newpage
% 

\end{document}